\documentclass[%
 aip,
 apl,
 amsmath,amssymb,
 reprint, 
 ]{revtex4-1}

\usepackage{graphicx}
\usepackage{dcolumn}
\usepackage{bm}
\usepackage[utf8]{inputenc}
\usepackage[T1]{fontenc}
\usepackage{mathptmx}
\usepackage{etoolbox}
\usepackage{physics}
\usepackage{color}
\usepackage[colorlinks=true,linkcolor=blue,citecolor=blue,urlcolor=blue]{hyperref}
\makeatletter
\def\@email#1#2{%
 \endgroup
 \patchcmd{\titleblock@produce}
  {\frontmatter@RRAPformat}
  {\frontmatter@RRAPformat{\produce@RRAP{*#1\href{mailto:#2}{#2}}}\frontmatter@RRAPformat}
  {}{}
}%
\makeatother
\begin{document}

\preprint{AIP/123-QED}

\title[Enhanced phase sensitivity of photon-recycled DSU(1,1) interferometer] {Enhancement in phase sensitivity in displacement-assisted SU(1,1) interferometer via photon recycling}
\author{Taj Kumar}
\affiliation{Department of Physics, Institute of Science, Banaras Hindu University, Varanasi-221005, Uttar Pradesh, India}
\author{Aviral Kumar Pandey}
\affiliation{Department of Physics, Institute of Science, Banaras Hindu University, Varanasi-221005, Uttar Pradesh, India}
\author{Anand Kumar}
\affiliation{Department of Physics, Institute of Science, Banaras Hindu University, Varanasi-221005, Uttar Pradesh, India}
\author{Devendra Kumar Mishra }%
\affiliation{Department of Physics, Institute of Science, Banaras Hindu University, Varanasi-221005, Uttar Pradesh, India} 
\email{kndmishra@gmail.com}

\date{\today}

\begin{abstract}
We propose a novel method for enhancing phase estimation in the displacement-assisted SU(1,1) (DSU(1,1)) interferometer by incorporating the photon recycling technique, evaluated under single-intensity detection (SID) and homodyne detection (HD) schemes. Our analysis showed that utilizing the photon recycling technique, the photon-recycled DSU(1,1) interferometer performs better than the conventional DSU(1,1) interferometer for some conditions. We also showed that this improvement is possible in both SID and HD schemes. In addition, to discuss the maximum sensitivity achieved by our proposed model, we have calculated the quantum Cram\'{e}r-Rao bound (QCRB) within the framework and found that our proposed model approaches the QCRB. Therefore, we believe that our findings offer a promising new approach to improve phase sensitivity through photon recycling.

\end{abstract}

\maketitle

%

\section{\label{Secton 1}Introduction}

Quantum metrology offers a robust and advanced framework to address the demand for high precision in  quantum information technologies, such as quantum sensors\cite{aslam2023quantum,brida2010experimental,genovese2016real,lawrie2019quantum}
and quantum imaging\cite{lemos2014quantum,proctor2018multiparameter,shih2007quantum,thomas2011real,hudelist2014quantum,li2018effects} and other quantum-enhanced applications \cite{michael2019squeezing,li2022quantum,li2024harnessing}. By using quantum features of light, quantum metrology achieves higher measurement precision than classical methods. In particular, quantum optical interferometers such as the Mach-Zehnder interferometer (MZI) \cite{zetie2000does,yadav2024quantum,shukla2021quantum,shukla2024quantum} and SU(1,1) interferometers \cite{yurke19862,ou2020quantum,marino2012effect,sharma2024enhancement} have become essential tools in quantum metrology. A conventional MZI consists of two beam splitters and its sensitivity can be significantly enhanced by replacing classical input light with non-classical states like twin Fock state \cite{campos2003optical}, two-mode squeezed state (TMSVS) \cite{anisimov2010quantum}, NOON state \cite{afek2010high,dowling2008quantum} and more recently, multiphoton catalyzed state \cite{zhang2021improved,zhao2024phase}. These quantum resources allow the system to exceed the shot-noise limit (SNL), defined as $1/\sqrt{\text{N}}$, \cite{beenakker1999photon,slussarenko2017unconditional}, and to approach the Heisenberg limit (HL), given by $1/\text{N}$ \cite{walmsley2015quantum, PhysRevA.85.042112}, where N is the average total photon number inside the interferometer. The ability to exceed the SNL and achieve the HL highlights the critical role of non-classical states in improving phase sensitivity and precision. Furthermore, significant advancements in phase sensitivity and noise reduction have been achieved by modifying the standard MZI setup. By replacing the passive beam splitters with active nonlinear elements, such as four-wave mixers \cite{slusher1985observation,gerry2000heisenberg} or optical parametric amplifiers (OPAs) \cite{leonhardt2003quantum, PhysRevA.78.012339,zuo2020quantum}, as first proposed by Yurke \cite{yurke19862}, the modified interferometers demonstrate a significant improvement in measurement precision. This modified interferometer is called an SU(1,1) interferometer. These active elements provide enhanced phase sensitivity by introducing quantum correlations and squeezing, thus reducing detection noise and pushing the system beyond classical performance limits. 
Building on these foundational concepts, Chang \textit{et al.} \cite{chang2022improvement} introduced an SU(1,1) interferometer design that incorporates a Kerr medium to induce a phase shift.  Despite photon losses, it significantly enhances the phase sensitivity and quantum Fisher information (QFI), a key metric for quantifying ultimate measurement precision in quantum systems.  Moreover, by applying non-Gaussian operations, like photon addition and subtraction within an SU(1,1) interferometer improves robustness against photon losses\cite{xin2021phase}. However, implementing these non-Gaussian techniques comes with considerable challenges, as they involve high operational complexity and significant costs \cite{zhang2021improved,zhang2021improving,carranza2012photon,xin2021phase}.

To address this challenge, local operations, such as the local displacement operator (LDO) \cite{PhysRevA.84.012335}, have emerged as promising solutions for enhancing phase sensitivity. Wei Ye \textit{et al.} \cite{ye2023quantum} introduced a SU(1,1) interferometer incorporating LDO, which is named a displacement-assisted SU(1,1) interferometer (DSU(1,1) interferometer). In addition to this approach, photon recycling has also gained attention as a powerful technique to further boost phase sensitivity. Several studies \cite{fritschel1992demonstration,meers1988recycling,mckenzie2004analysis, Li:23} have demonstrated that photon recycling can significantly enhance signal intensity by increasing the mean photon number circulating within the system, as evidenced in the case of the Michelson interferometer. Notably, Li Dong \textit{et al.} \cite{Li:23} proposed a quantum-enhanced phase estimation method by integrating photon recycling into the MZI, achieving a remarkable 9.32-fold improvement in phase sensitivity.

In this work, we present a novel approach to improving phase estimation in the DSU(1,1) interferometer by integrating photon recycling techniques. We conduct a thorough analysis of the quantum Cramér-Rao bound (QCRB) and phase sensitivity of the photon-recycled DSU(1,1) (PR-DSU(1,1)) interferometer and compare its performance against the SNL. Our results demonstrate that this modified configuration significantly enhances phase sensitivity and effectively surpasses the SNL. This highlights photon recycling as a promising strategy for achieving improved phase sensitivity in quantum interferometric systems.

The structure of the paper is as follows: In Section \ref{Section 2}, we provide an overview of the key concepts related to the standard DSU(1,1) interferometer and introduce our proposed scheme. Section \ref{Section 3} covers quantum phase estimation using both single-intensity detection and homodyne detection schemes. In Section \ref{Section 4}, we examine the QCRB of the photon-recycled DSU(1,1) interferometer under various parameter settings. In Section \ref{Section 5}, we have compared the performance of the PR-DSU(1,1) interferometer with the conventional DSU(1,1) interferometer. In Section \ref{Section 6}, we compare the SNL with phase sensitivities under both detection schemes as well as QCRB  of the PR-DSU(1,1) interferometer. Finally, Section \ref{Section 7} summarizes our findings.

\section{PR-DSU(1,1) interferometer model}\label{Section 2}
\begin{figure}[htbp]
\centering\includegraphics[width=3.45in, height=2.8in]{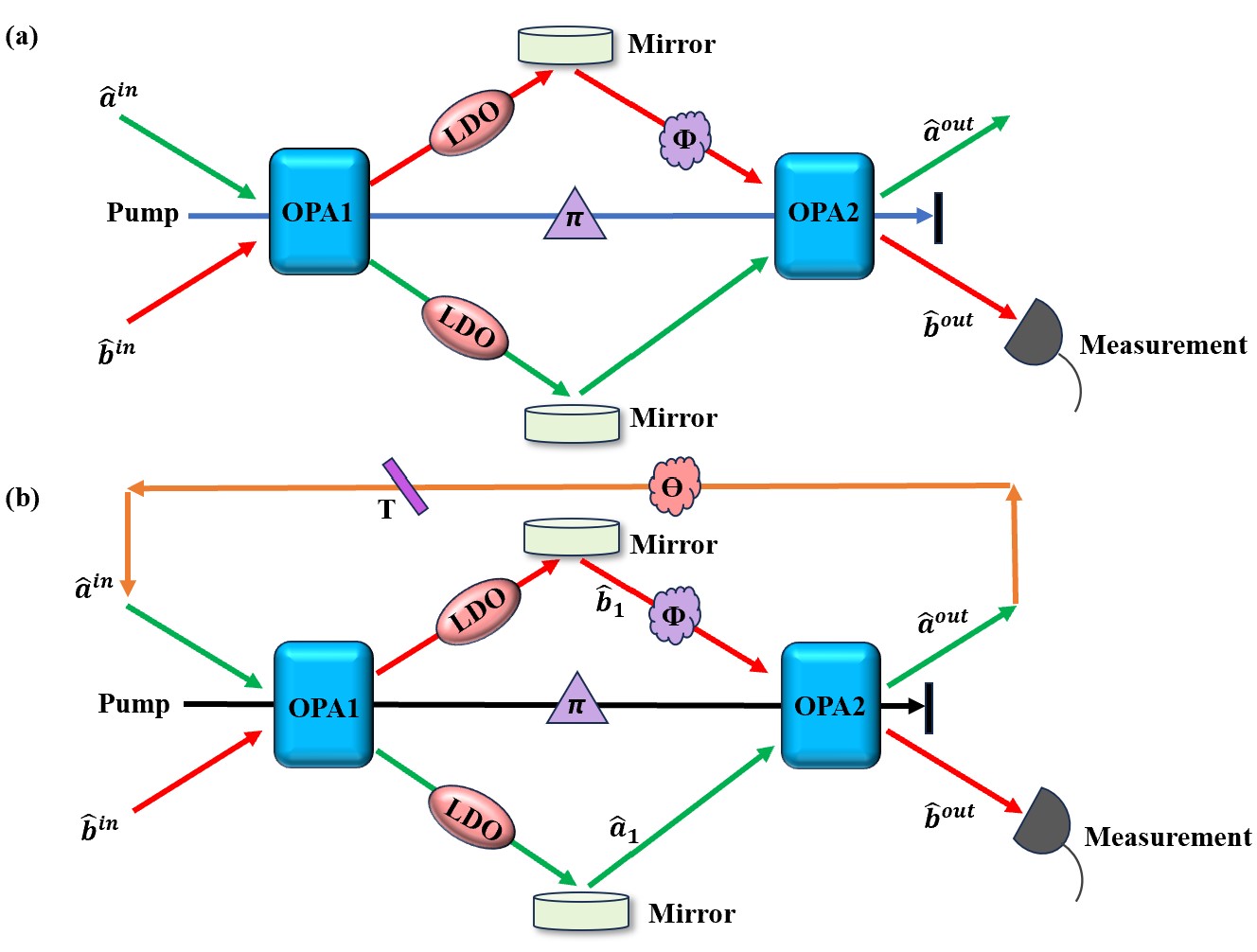}
\caption{\label{Fig._1} Scheme for phase estimation with photon recycling technique:
(a) Schematic diagram of a DSU(1,1) interferometer with the output beam "a" disregarded. 
(b) The modified scheme with the output mode "a" re-injected into the input mode "a" (PR-DSU(1,1) interferometer). We are assuming that the output beam experiences a phase shift $\theta$ and photon loss of $\sqrt{1-\text{T}}$ before re-injection.}
\end{figure}
As illustrated in Fig. 1(a), the DSU(1,1) interferometer consists of two OPAs and two LDOs with a phase shift $\phi$ in arm ``b''.  In quantum metrology, the OPA process is generally described by the unitary operator $\hat{\text{U}}_{\text{OPA}}=e^{-s\hat{a}^{\dagger}\hat{b}^{\dagger}+s^{*}\hat{a}\hat{b}}$, where $s=g e^{i \eta}$ is the two-mode squeezing parameter with the OPA gain g and phase information $\eta$ and $\hat{a}(\hat{a}^{\dagger})$, $\hat{b}(\hat{b}^{\dagger})$ are the annihilation (creation) operators of modes ``a'' and ``b'', respectively. After the input states pass through the OPA1, both paths undergo the same LDO process, denoted as $\hat{D}_{a}(\gamma)=e^{\gamma \hat{a}^{\dagger}-\gamma^{*} \hat{a}}$ and $\hat{D}_{b}(\gamma)=e^{\gamma \hat{b}^{\dagger}-\gamma^{*} \hat{b}}$, where $\gamma=|\gamma|e^{i \delta_{\gamma}}$.
The phase shift process in arm ``b'' can be described as $\hat{\text{U}}_{\phi}=e^{i \phi\hat{b}^{\dagger}\hat{b}}$.

The relationship between the input and output modes in the DSU(1,1) interferometer can be described as follows \cite{ye2023quantum}

\begin{equation}\label{Eq. 1}
\begin{aligned}
\hat{a}^{\text{out}} & =W_{1}+Y\hat{a}^{\text{in}}-Z\hat{b}^{\dagger}_{\text{in}}, \\
\hat{b}^{\text{out}} & =W_{2}+e^{i \phi}(Y\hat{b}^{\text{in}}-Z\hat{a}^{\dagger}_{\text{in}}),
\end{aligned}
\end{equation}
where $W_{1}$ and $W_{2}$ ocuured due to LDO process, and 

\begin{equation}\label{Eq. 2}
\begin{aligned}
Y & =\cosh{\text{g}_{1}}\cosh{\text{g}_{2}}+e^{i (\eta_{2}-\eta_{1}-\phi)}\sinh{\text{g}_{1}}\sinh{\text{g}_{2}}, \\
Z & =e^{i \eta_{1}}\sinh{\text{g}_{1}}\cosh{\text{g}_{2}}+e^{i (\eta_{2}-\phi)}\cosh{\text{g}_{1}}\sinh{\text{g}_{2}}, \\
W_{1} & =\gamma \cosh{\text{g}_{2}} - \gamma^{*} e^{i (\eta_{2}-\phi)}\sinh{\text{g}_{2}}, \\
W_{2} & =\gamma e^{i \phi}  \cosh{\text{g}_{2}} - \gamma^{*} e^{i \eta_{2}}\sinh{\text{g}_{2}}.
\end{aligned}
\end{equation}
\\
Here, $\text{g}_{1}(\text{g}_{2})$ and $\eta_{1}(\eta_{2})$ denote the gain and phase shift for OPA1(OPA2), respectively.

We propose a novel approach where the output beam from mode "a" is re-injected into input port "a", as shown in Fig. 1(b). However, in practical scenarios, photon losses are unavoidable during light propagation in the recycling arm. To account for this, we analyze the effects of photon loss during propagation, modeling it with a fictitious beam splitter characterized by a transmission coefficient, T. Furthermore, we assume that the re-injected beam experiences a phase shift $\theta$ before it is fed back into input port "a". 

\begin{figure}[htbp]
\centering\includegraphics[width=1\linewidth,height=1.4in]{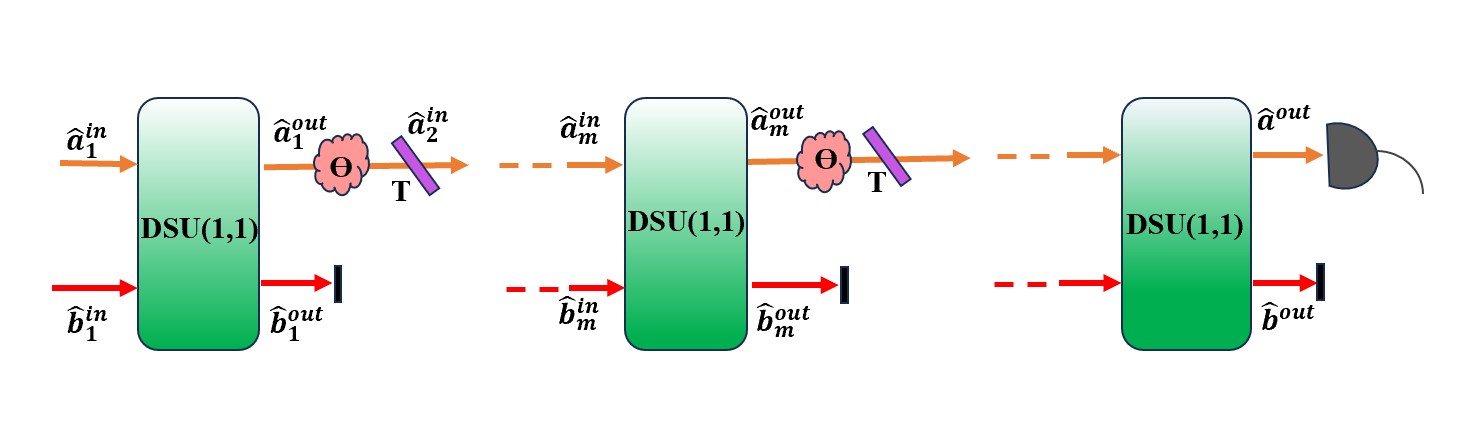}
\caption{\label{Fig._2.}The iterative-structure of series of DSU(1,1) interferometers where $(k-1)^{th}$ output mode ``a'' is injected into $k^{th}$ input mode ``a''. This model is equivalent to the scheme in Fig. 1(b). }
\end{figure}

To describe the input-output relationship in the photon-recycled DSU(1,1)(PR-DSU(1,1)) interferometer, we introduce an iterative method, as shown in Fig.~\ref{Fig._2.}.
The setup shown in Fig. 1(b) is equivalent to a sequence of conventional DSU(1,1) interferometers arranged in an iterative manner, where each output mode ``a'' is fed into the next input port 
``a''. This iterative optical system establishes the input-output relation in the case of PR-DSU(1,1) interferometer as below
\begin{widetext}
     \begin{equation} \label{Iterative transformation}
\begin{array}{c}
\begin{pmatrix}\hat{a}_{1}^{\text{out}} \\\hat{b}_{1}^{\dagger \text{out}}\end{pmatrix} = \hat{U}_{\text{DSU}} \begin{pmatrix}\hat{a}_{1}^{\text{in}} \\\hat{b}_{1}^{\dagger \text{in}}\end{pmatrix},
\begin{pmatrix}\hat{a}_{2}^{\text{out}} \\\hat{b}_{2}^{\dagger \text{out}}\end{pmatrix} = \hat{U}_{\text{DSU}} \begin{pmatrix}\hat{a}_{2}^{\text{in}} \\\hat{b}_{1}^{\dagger \text{in}}\end{pmatrix}, \cdots, 
\begin{pmatrix}\hat{a}_{m}^{\text{out}} \\\hat{b}_{m}^{\dagger \text{out}}\end{pmatrix} = \hat{U}_{\text{DSU}} \begin{pmatrix}\hat{a}_{m}^{\text{in}} \\\hat{b}_{1}^{\dagger \text{in}}\end{pmatrix},\cdots
\end{array}
\end{equation}
\end{widetext}
where
\begin{equation}\label{Recycled mode a and b}
\begin{aligned}
\centering
\hat{a}_{k}^{\text{in}}& =\sqrt{\text{T}} e^{i \theta} \hat{a}_{k-1}^{\text{out}} + \sqrt{1-\text{T}}\hat{v}_{a}, \\
\centering
\hat{b}_{k}^{\dagger \text{in}} & =\hat{b}_{1}^{\dagger \text{in}},
\end{aligned}
\end{equation}
with $k=2,3,...,m$.

Based on Eqs. (\ref{Iterative transformation}) and (\ref{Recycled mode a and b}), the input-output relation for $m^{th}$ DSU(1,1) interferometer is given by
\begin{widetext}
    \begin{equation}\label{Eq. 5}
\begin{aligned}
\hat{a}_{m}^{\text{out}} &= W_{1} - Z \hat{b}_{1}^{\dagger \text{in}} + Y A^{m-1} \hat{a}_{1}^{\text{in}} + Y \frac{1-A^{m-1}}{1-A} B, \\
\hat{b}_{m}^{\dagger \text{out}} &= W_{2}^{*} + Y^{*} e^{-i \phi} \hat{b}_{1}^{\dagger \text{in}} - Z^{*} e^{-i \phi} A^{m-1} \hat{a}_{1}^{\text{in}} - Z^{*} e^{-i \phi} \frac{1-A^{m-1}}{1-A} B,
\end{aligned}
\end{equation}
\end{widetext}

where
\begin{equation}\label{Eq. 6}
\begin{aligned}
A &=Y\sqrt{\text{T}}e^{i \theta}, \\
B &=\sqrt{\text{T}}e^{i \theta}(W_{1}-Z\hat{b}_{1}^{\dagger in})+\sqrt{1-\text{T}}\hat{v}_{a}.
\end{aligned}
\end{equation}
Now, in the limit m tends to infinity $(m \to\infty)$, we get the relationship between the input and output modes of the PR-DSU(1,1) interferometer as below

\begin{equation}\label{input-output relation of PR-DSU}
\begin{aligned}
 \hat{a}^{\text{out}} & =C_{1} (W_{1}-Z\hat{b}_{1}^{\dagger \text{in}}+Y \sqrt{1-\text{T}}\hat{v}_{a}), \\
\hat{b}^{\dagger \text{out}} & =C_{2}+ C_{3} e^{-i\phi}\hat{b}_{1}^{\dagger \text{in}}-C_{1}Z^{*}\sqrt{1-\text{T}}e^{-i \phi}\hat{v}_{a}.
\end{aligned}
\end{equation}
where
\begin{equation}\label{Constants}
\begin{aligned}
 C_{1} & =\left(1+\frac{Y\sqrt{\text{T}}}{e^{-i \theta}-Y\sqrt{\text{T}}}\right), \\
C_{2} & =\left(W_{2}^{*}-\frac{Z^{*}\sqrt{\text{T}}W_{1} e^{-i\phi}}{e^{-i \theta}-Y\sqrt{\text{T}}}\right),\\
C_{3} & =\left(Y^{*}+\frac{|Z|^{2}\sqrt{\text{T}}}{e^{-i \theta}-Y\sqrt{\text{T}}}\right).
\end{aligned}
\end{equation}
On putting $\text{T}=0$ in Eq. \ref{input-output relation of PR-DSU}, these output field modes reduce to those of the conventional DSU(1,1) interferometer, given by Eq. \ref{Eq. 1}.
\section{Phase sensitivity of PR-DSU(1,1) interferometer}\label{Section 3}

We have already provided a detailed description of the framework for the PR-DSU(1,1) interferometer. In the following sections, we assume a balanced configuration for the PR-DSU(1,1) interferometer, where $\eta_{2}-\eta_{1}=\pi$ and $\text{g}_{2}=\text{g}_{1}=\text{g}$ (where $\eta_{1}=0$ and $\eta_{2}=\pi$), without loss of generality. The output signal is measured at the end of the interferometer, allowing us to determine the sensitivity of the estimated parameters. We are considering the input state $\ket{\psi_{in}}=\ket{0}_{a}\otimes\ket{\xi}_{b}$, where $\ket{0}_{a}$ is the vacuum state at port "a"  and $\ket{\xi}_{b}=\hat{S}(\xi)\ket{0}_{b}$ (with the squeezing operator $\hat{S}(\xi)=e^{(\xi^{*}\hat{b}^{2}-\xi\hat{b}^{\dagger 2})/2}$ ($\xi=\text{r} e^{i\delta_{\xi}})$) is the squeezed vacuum state at port "b". For the sake of discussion, we set $\delta_{\xi}=0$. Similarly, we 
 take $\delta{\gamma}=0$.

 Quantum metrology offers several detection methods, including single-intensity detection \cite{yadav2024quantum}, homodyne detection \cite{ye2023quantum,sharma2024enhancement}, and parity detection \cite{shukla2023improvement,sharma2024super}. In this work, we will evaluate the phase sensitivity of the interferometer under the simpler single-intensity detection (SID) scheme.  Additionally, we apply a homodyne detection (HD) scheme to analyze the sensitivity of the PR-DSU(1,1) interferometer for comparison, as it offers higher sensitivity by effectively monitoring the quadrature components of light. In the following subsections, we will provide the simplified expressions for phase-sensitivity under SID, and HD schemes along with the total photon number inside the PR-DSU(1,1) interferometer.
 
\subsection{Single intensity detection (SID) scheme}
To obtain information about the parameter $\phi$, we employ the SID scheme in output mode ``b'', as shown in Fig. \ref{Fig._1}(b). The photon number operator at output port ``b'' is mathematically defined as follows

\begin{equation}\label{Eq. 9}
\hat{\text{N}}_{\text{b}}=\hat{\text{b}}^{\text{out}\dagger }\hat{\text{b}}^{\text{out}}.
\end{equation}
\\
Using the linear error propagation formula, the phase sensitivity in our scheme is expressed as follows
\begin{equation}\label{Eq. 10}
(\Delta \phi^\text{PR})_{\text{SID}}=\frac{\Delta\hat{\text{N}}_{\text{b}}}{\left|\frac{\partial\left\langle\hat{\text{N}}_{\text{b}}\right\rangle}{\partial \phi}\right|}.
\end{equation}
where $\Delta\hat{\text{N}}_{\text{b}}$= $\sqrt{\left\langle\hat{\text{N}}_{\text{b}}^{2}\right\rangle-\left\langle\hat{\text{N}}_{\text{b}}\right\rangle^{2}}.$ Here, $\left\langle ... \right\rangle$ is the expectation value of the operator at port ``b'' with respect to the input state $\ket{\psi_{\text{in}}}$. In this setup, using Eq. \ref{input-output relation of PR-DSU}, the average value of photon number, $\left\langle\hat{\text{N}}_{\text{b}}\right\rangle$ and its standard deviation $\Delta\hat{\text{N}}_{\text{b}}$ at the output port ``b'' are given by
\begin{equation}\label{Eq. 11}
\left\langle\hat{\text{N}}_{\text{b}}\right\rangle=|C_{2}|^2+|C_{3}|^2 \sinh(r)^2+|C_{1}|^2 |Z|^2 (1-\text{T}).
\end{equation}
\begin{widetext}
\begin{equation}\label{Eq. 12}
\begin{split}
\Delta\hat{\text{N}}_{\text{b}}=\sqrt{|C_{2}|^2|C_{3}|^2 \cosh(2\text{r}) + |C_{1}|^2 |Z|^2 (1-\text{T})(|C_{2}|^2+|C_{3}|^2)-\sinh{(2\text{r})} \text{Re}[C_{3}^{*2}C_{2}^{2} e^{2i \phi}]+2\cosh^2(\text{r})\sinh^2(\text{r})|C_{3}|^{4}}.
\end{split}
\end{equation}
\end{widetext}
Putting the values from Eqs. (\ref{Eq. 11}) and (\ref{Eq. 12}) into Eq. (\ref{Eq. 10}), we finally get the expression of phase sensitivity of the PR-DSU(1,1) interferometer under the SID scheme.
\subsection{Homodyne detection (HD) scheme}
To obtain information about the parameter $\phi$, we employ the HD scheme to determine the quadrature of the output mode ``b'' as shown in Fig. 1(b). The output quadrature at port ``b'' is mathematically defined as follows
\begin{equation}\label{Eq. 13}
\hat{X}=\frac{\hat{b}^{\text{out}\dagger }+\hat{b}^{\text{out}}}{\sqrt{2}}.
\end{equation}
Using the linear error propagation formula, the phase sensitivity in our scheme is expressed as follows
\begin{equation}\label{Eq. 14}
(\Delta \phi^\text{PR})_{\text{HD}}=\frac{\Delta\hat{X}}{\left|\frac{\partial\left\langle\hat{X}\right\rangle}{\partial \phi}\right|},
\end{equation}
where $\Delta\hat{X}$= $\sqrt{\left\langle\hat{X}^{2}\right\rangle-\left\langle\hat{X}\right\rangle^{2}}.$ Here, $\left\langle ... \right\rangle$ is the expectation value of the operator at output port ``b'' with respect to the input state $\ket{\psi_{in}}$. In this setup, the average value of quadrature operator $\left\langle\hat{X}\right\rangle$ and its standard deviation  $\Delta\hat{X}$ at output port ``b'' are given by

\begin{equation}\label{Eq. 15}
\left\langle\hat{X}\right\rangle=\sqrt{2} \text{Re}[C_{2}],
\end{equation}
\begin{widetext}
    \begin{equation}\label{Eq. 16}
\begin{split}
\Delta \hat{X} =\sqrt{\frac{|C_{3}|^2 \cosh(2\text{r}) + |C_{1}|^2 |Z|^2 (1-\text{T})-\sinh{(2\text{r})} \text{Re}[C_{3}^{2} e^{-2i \phi}]}{2}}.
\end{split}
\end{equation}
\end{widetext}
Putting the values from Eqs. (\ref{Eq. 15}) and (\ref{Eq. 16}) into Eq. (\ref{Eq. 14}), we get the phase sensitivity of PR-DSU(1,1) interferometer under homodyne detection.

\subsection{Total photon number inside the PR-DSU(1,1) interferometer}
It is well established that the phase sensitivity of an interferometer depends on the total mean photon number $N_{\text{total}}$ within the interferometer, which accounts for the photons interacting with and sensing the phase shift. In our scheme, it is given by

\begin{equation}\label{NT}
N_{\text{total}}=\bra{\psi_{\gamma}}(\hat{a}_{1}^{\dagger}\hat{a}_{1}+\hat{b}_{1}^{\dagger}\hat{b}_{1})\ket{\psi_{\gamma}},
\end{equation}
where $\ket{\psi_{\gamma}}=\hat{D}_{a}(\gamma)\hat{D}_{b}(\gamma)\hat{U}_{\text{OPA1}}\ket{\psi_{\text{in}}}$ is the probe state just after the LDO. 
In the case of PR-DSU(1,1) interferometer, we have
\begin{equation} \label{transformation}
\begin{aligned}
\begin{pmatrix}
\hat{a}_{1} \\
\hat{b}_{1}^{\dagger}
\end{pmatrix}
&= \hat{D}_{a}(\gamma)\hat{D}_{b}(\gamma)\hat{U}_{\text{OPA1}}
\begin{pmatrix}
\hat{a}^{\text{in}} \\
\hat{b}^{\dagger \text{in}}
\end{pmatrix}, \\
\hat{a}^{\text{in}} &= \sqrt{\text{T}} e^{i \theta} \hat{a}^{\text{out}} + \sqrt{1-\text{T}} \hat{v}_{a}.
\end{aligned}
\end{equation}
So, using Eqs. (\ref{NT}) and (\ref{transformation}) with our input state $\ket{\psi_{\text{in}}}$, we get
\begin{widetext}
\begin{equation}\label{NTv}
\begin{split}
    N_{\text{total}}& =2|\gamma|^{2}+\text{T}|C_{1}|^{2}\Bigg(\cosh{(2\text{g})}\Big(|W_{1}|^{2}+|Z|^{2}\cosh^2{(\text{r})}\Big)-|Z|^{2}\sinh^2{(\text{g})}\Bigg) +(\sinh^2{(\text{g})}\\
    & +\cosh{(2\text{g})}\sinh^2{(\text{r})})+\sinh^2{(\text{g})}(1-\text{T})(1+|Y|^{2}+2\sqrt{\text{T}}\text{Re}[Y e^{i \theta}])+\sqrt{\text{T}} \sinh{(2\text{g})}\\
    & \cosh{(2\text{r})} \text{Re}[ZC_{1}e^{i \theta}]
    +2|\gamma|\sqrt{\text{T}}\Big(\cosh{(\text{g})}-\sinh{(\text{g})}\Big) \text{Re}[C_{1}W_{1}e^{i \theta}].
\end{split}
\end{equation}
\end{widetext}
Hence, the SNL of our system will be
 
\begin{equation}\label{SNL}
\Delta \phi_{\text{SNL}}=\frac{1}{\sqrt{N_{\text{total}}}}.
\end{equation}

\section{Quantum Cram\'{e}r-Rao bound of PR-DSU(1,1) interferometer}\label{Section 4}
Quantum Fisher information (QFI) and hence the quantum Cram\'{e}r-Rao bound (QCRB) \cite{helstrom1967minimum,helstrom1969quantum,braunstein1994statistical,ataman2020single,ataman2022quantum} theoretically define the ultimate measurement precision and do not depend upon the detection scheme. By optimizing experimental settings and utilizing quantum resources, it is possible to approach this theoretical limit, thereby improving the accuracy and precision of phase measurements. In our scenario, the output mode b always remains a pure Gaussian state as the input light is a squeezed vacuum state, which is a pure Gaussian state, and since OPA and LDO act as Gaussian operations, the output still remains the pure Gaussian state propagated through the OPA and LDO (including the fictitious beam splitter for photon loss, which also acts as a Gaussian operation \cite{Li:23}). 
Now, the QFI for an arbitrary pure state, assuming ideal conditions in our scheme, can be stated as\cite{ataman2020single,ataman2022quantum}
\begin{equation}\label{FQ}
F_{Q} = 4\Bigg[\bra{\psi_{\phi}^{'}}\ket{\psi_{\phi}^{'}}-\Big|\bra{\psi_{\phi}^{'}}\ket{\psi_{\phi}}\Big|^{2}\Bigg],
\end{equation}
where $\ket{\psi_{\phi}}=e^{i\phi\hat{b}^{\dagger}_{1}\hat{b}_{1}}\ket{\psi_{\gamma}}$
is the state prior to the second OPA and $\ket{\psi_{\phi}^{'}}=\partial\ket{\psi_{\phi}}/\partial\phi$.
Thus, the QFI can be calculated as
\begin{equation}\label{Fq}
F_{Q} = 4\Big[\bra{\psi_{\gamma}}({\hat{b}_{1}}^{\dagger}\hat{b}_{1})^{2}\ket{\psi_{\gamma}}-\Big|\bra{\psi_{\gamma}}\hat{b}^{\dagger}_{1}\hat{b}_{1}\ket{\psi_{\gamma}}\Big|^{2}\Big].
\end{equation}

Using Eq. \ref{Fq} with Eq. \ref{transformation}, QCRB of the PR-DSU(1,1) interferometer can be obtained as
\begin{equation}\label{QCRB}
\Delta\phi_{\text{QCRB}}^{\text{PR}}=\frac{1}{\sqrt{\eta F_{Q}}},
\end{equation}
where $\eta$ is the number of trials (for simplicity, we take $\eta=1$). According to Eq. \ref{QCRB}, a higher value of $F_{Q}$ results in smaller $\Delta\phi_{\text{QCRB}}$, indicating improved phase sensitivity.

\section{Comparison between performances of conventional DSU(1,1) and PR-DSU(1,1) intereferometers} \label{Section 5}
\subsection{Phase sensitivity}
Following the same steps in Section \ref{Section 3} with input state $\ket{\psi_{in}}$ and expression for output modes of the DSU(1,1) interferometer, given in Eq. \ref{Eq. 1}, we obtain the average value of photon number, $\left\langle\hat{\text{N}}_{\text{b}}\right\rangle$ and its standard deviation $\Delta\hat{\text{N}}_{\text{b}}$ at the output port ``b'' of the conventional DSU(1,1) interferometer are given by
\begin{equation}\label{Eq. 24}
\left\langle\hat{\text{N}}_{\text{b}}\right\rangle=|W_{2}|^2+|Y|^2 \sinh(\text{r})^2+ |Z|^2.
\end{equation}
\begin{widetext}
\begin{equation}\label{Eq. 25}
\begin{split}
\Delta\hat{\text{N}}_{\text{b}}=\sqrt{|W_{2}|^2|Y|^2 \cosh(2\text{r}) + |Z|^2 (|W_{2}|^2+|Y|^2)-\sinh{(2\text{r})} \text{Re}[Y^{2}W_{2}^{*2} e^{2i \phi}]+2\cosh^2(\text{r})\sinh^2(\text{r})|Y|^{4}}.
\end{split}
\end{equation}
\end{widetext}
Putting the values from Eqs. (\ref{Eq. 24}) and (\ref{Eq. 25}) into the standard error propagation formula, we finally get the expression of the phase sensitivity of the 
conventional DSU(1,1) interferometer under the SID scheme, $(\Delta \phi^{\text{Conv.}})_{\text{SID}}$.

Similarly, we obtain the average value of quadrature operator $\left\langle\hat{X}\right\rangle$ and its standard deviation  $\Delta\hat{X}$ at output port ``b'' of the conventional DSU(1,1) interferometer as below
\begin{equation}\label{Eq. 26}
\left\langle\hat{X}\right\rangle=\sqrt{2} \text{Re}[W_{2}],
\end{equation}
    \begin{equation}\label{Eq. 27}
\begin{split}
\Delta \hat{X} =\sqrt{\frac{|Y|^2 \cosh(2\text{r}) +  |Z|^2 -\sinh{(2\text{r})} \text{Re}[Y^{*2} e^{-2i \phi}]}{2}}.
\end{split}
\end{equation}
Putting the values from Eqs. (\ref{Eq. 26}) and (\ref{Eq. 27}) into the standard error propagation formula, we get the phase sensitivity of conventional DSU(1,1) interferometer under homodyne detection $(\Delta \phi^{\text{Conv.}})_{\text{HD}}$

In order to show the improvement in phase-sensitivity of PR-DSU(1,1) over the conventional DSU(1,1) interferometer, we define
\begin{equation}\label{Eq. 28}
\Sigma=\frac{\Delta \phi^{\text{Conv.}}}{\Delta\phi^{\text{PR}}},
\end{equation}
where $\Sigma$ represents the enhancement factor for phase sensitivity in our scheme compared to the conventional DSU(1,1) interferometer.
\\
\begin{figure}[htbp]
\centering
\includegraphics[width=1\linewidth,height=2.3in]{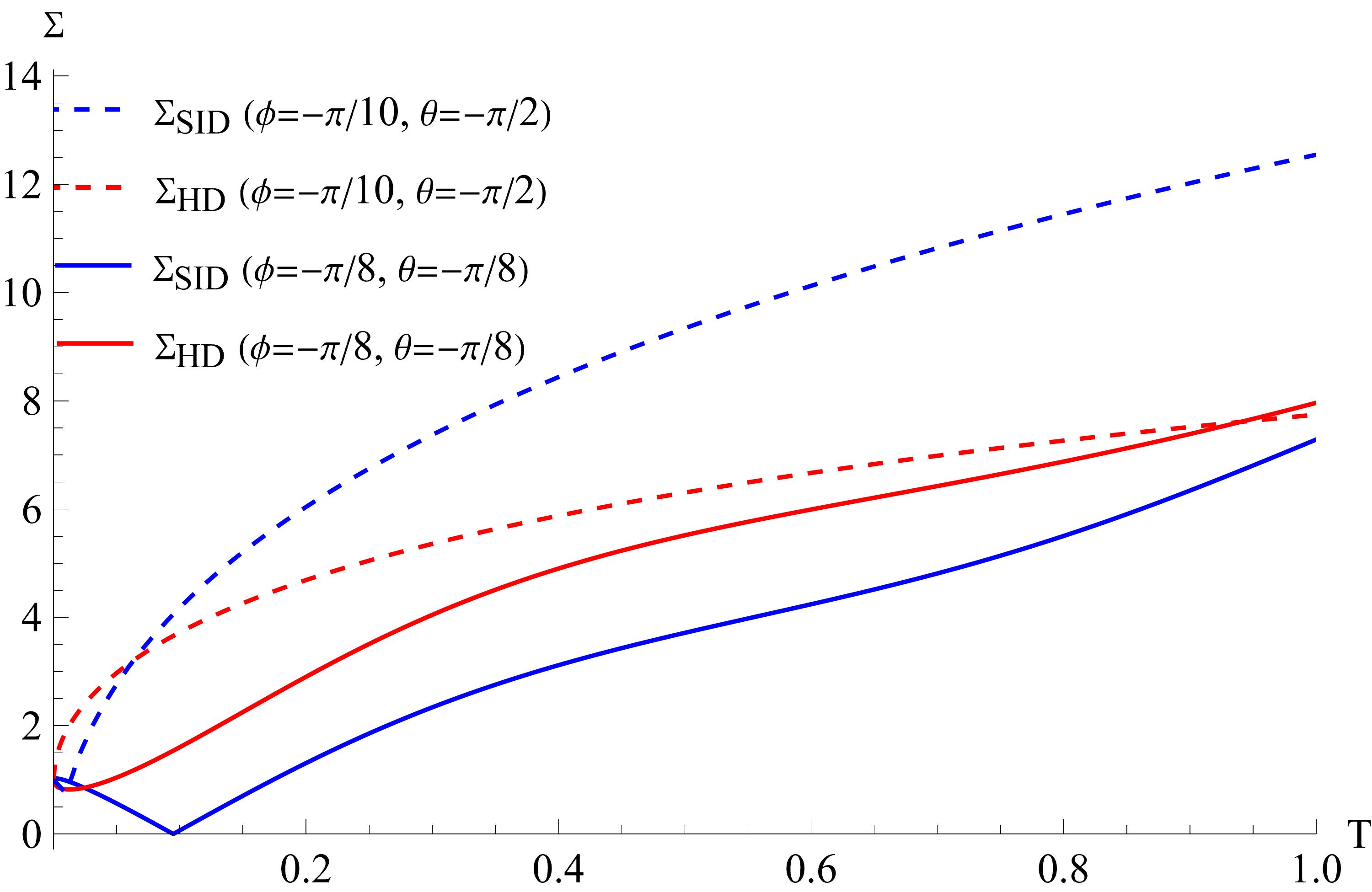}
\caption{\label{Fig._3} The enhancement factor $\Sigma$ as a function of T. Other parameters are $\text{g}=1$, $\text{r}=0.5$, $|\gamma|=1$.}
\end{figure}
Next, we examine the effect of T on the enhancement factors $\Sigma$ under both detection schemes, as shown in Fig. ~\ref{Fig._3}. 
We observe that depending upon the values of $\theta$ and $\phi$, we can achieve the enhancement factor $\Sigma$ greater than unity,  clearly indicating the improvement in both phase sensitivity of PR-DSU(1,1) interferometer as compared to conventional DSU(1,1) interferometer. Moreover, we observe that $\Sigma$ increases with increasing T, i.e., the increasing number of recycled photons, highlighting the importance of photon recycling.
\subsection{QCRB}
Following the same steps as in Section \ref{Section 4} and using Eq. \ref{transformation}, by considering  $\hat{a}^{\text{in}}$ to be annihilation operator for vacuum at input mode a with $\text{T}=0$, we obtain the expression for QCRB of the DSU(1,1) interferometer, denoted as $\Delta\phi_{\text{QCRB}}^{\text{Conv.}}$.

In order to show the improvement in QCRB of PR-DSU(1,1) over the DSU(1,1) interferometer, we define
\begin{equation}\label{Eq. 29}
\Xi=\frac{\Delta\phi_{\text{QCRB}}^{\text{Conv.}}}{\Delta\phi_{\text{QCRB}}^{\text{PR}}},
\end{equation}
where $\Xi$ represents the enhancement factor of QCRB in our scheme compared to the conventional DSU(1,1) interferometer.
\begin{figure}[htbp]
\centering
\includegraphics[width=1\linewidth]{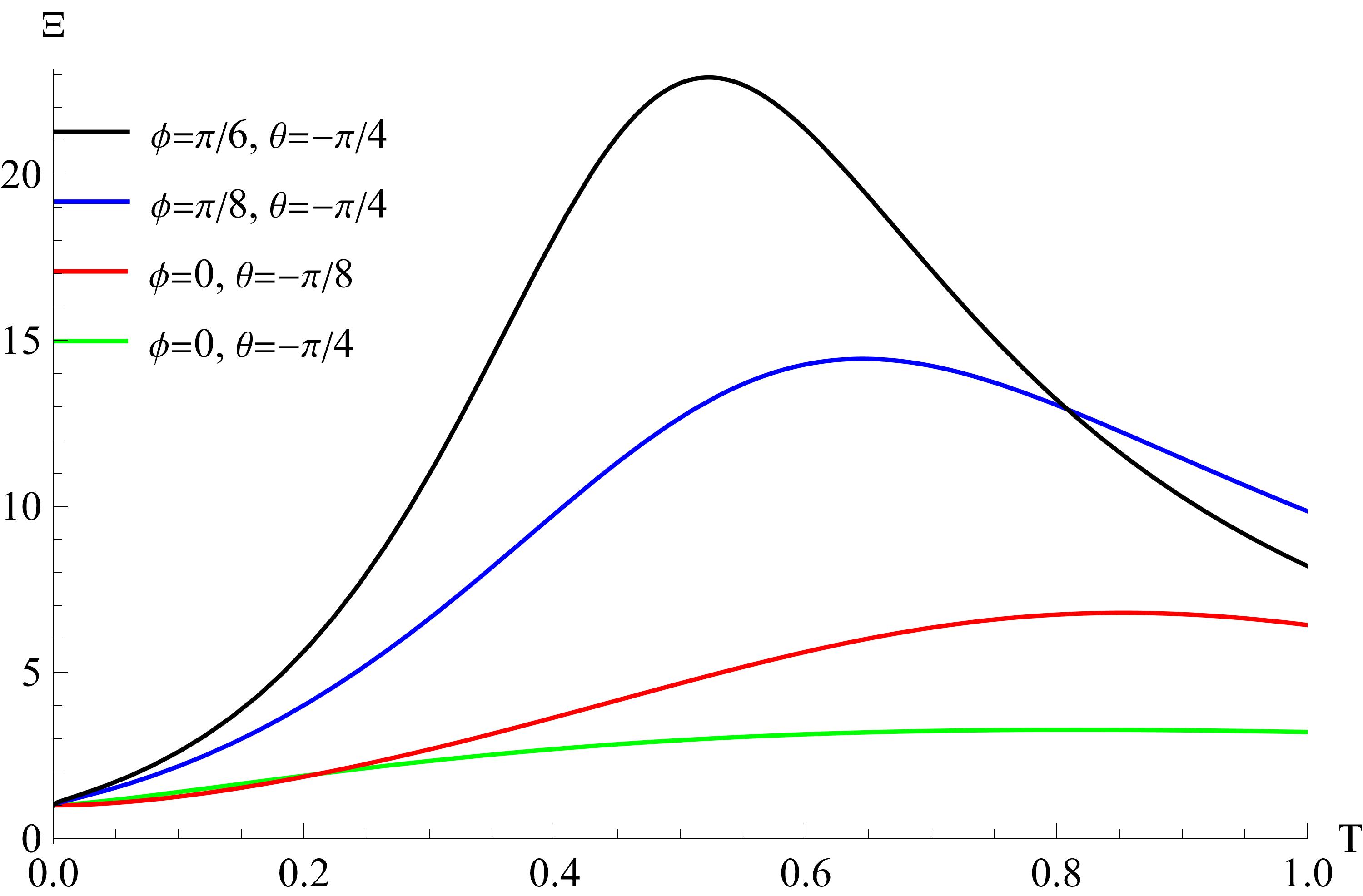}
\caption{\label{Fig._4} The enhancement factor $\Xi$ as a function of T. Other parameters are $\text{g}=1$, $\text{r}=0.5$, $|\gamma|=1$.}
\end{figure}

{Next, we examine the effect of T on the enhancement factors $\Xi$ as shown in Fig. ~\ref{Fig._4}. 
We again observe that depending upon the values of $\theta$ and $\phi$, we can achieve this enhancement factor $\Xi$ greater than unity along with significant improvement in a certain range of T,  clearly indicating the improvement in the QCRB of PR-DSU(1,1) interferometer as compared to conventional DSU(1,1) interferometer,  further highlighting the importance of photon recycling.}

\section{Improvement in the performance of PR-DSU(1,1) interferometer relative to shot-noise limit (SNL)} \label{Section 6}

\subsection{{Phase sensitivity}}
In order to show the improvement in phase sensitivity over SNL, we define
\begin{equation}\label{Eq. 30}
\Gamma=\frac{\Delta \phi_{\text{SNL}}}{\Delta\phi^{\text{PR}}},
\end{equation}
where $\Gamma$ represents the enhancement factor of phase sensitivity in our scheme compared to the SNL.
Next, we will present and explain our findings on improvement in phase sensitivity across different scenarios by conducting a detailed graphical analysis, as outlined below.

\begin{figure}[htbp]
\centering
\includegraphics[width=1\linewidth, height=1.2in]{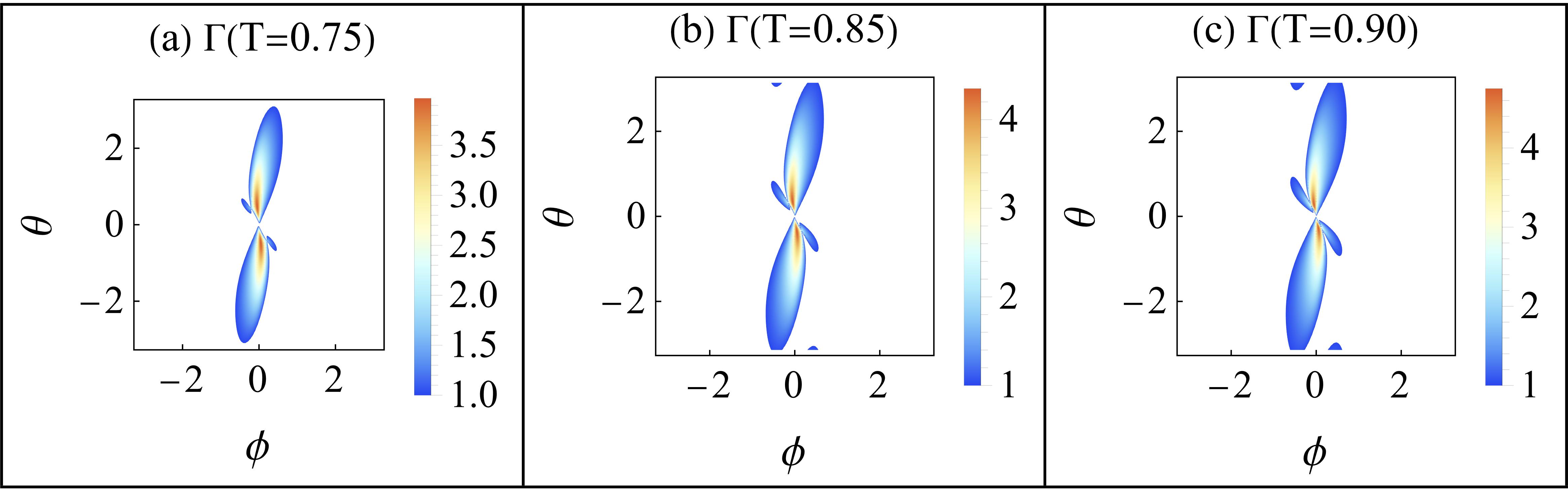}
\par\medskip 
\text{(A)}
\vskip\baselineskip 
\includegraphics[width=1\linewidth, height=1.2in]{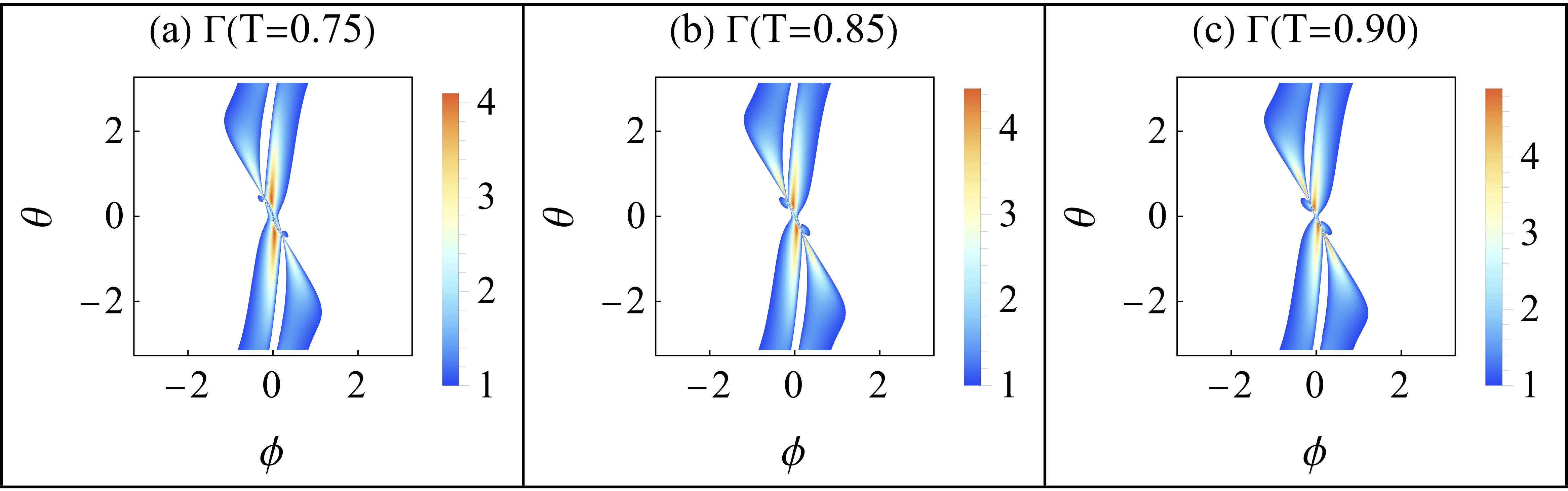}
\par\medskip 
\text{(B)}
\caption{The enhancement factor $\Gamma$, plotted as a function of phase angles $\theta$ and $\phi$ for (A) single-intensity detection and (B) homodyne detection. The plots are shown for transmission coefficients (a) $\text{T}=0.75$, (b) $\text{T}=0.85$, and (c) $\text{T}=0.95$, with the remaining parameters fixed at $\text{g}=1.2$, $\text{r}=0.5$, and $|\gamma|=1$.}
\label{Fig. 5}
\end{figure}

To illustrate the relationship of enhancement factor $\Gamma$ with phases $\theta$ and $\phi$ for different transmission coefficient T, we refer to Figs. ~\ref{Fig. 5}(A) (SID scheme) and ~\ref{Fig. 5}(B) (HD scheme). We observe that in both detection schemes, in the presence of moderate photon loss (up to 30\% loss), the enhancement factor $\Gamma$ can reach above unity (as shown in the colored region), indicating that our scheme achieves phase sensitivity beyond the SNL. We also see that the enhancement factor decreases with decreasing transmission coefficient (increasing photon loss) but remains above unity.

\begin{figure}[htbp]
\centering
\includegraphics[width=1\linewidth, height=1.2in]{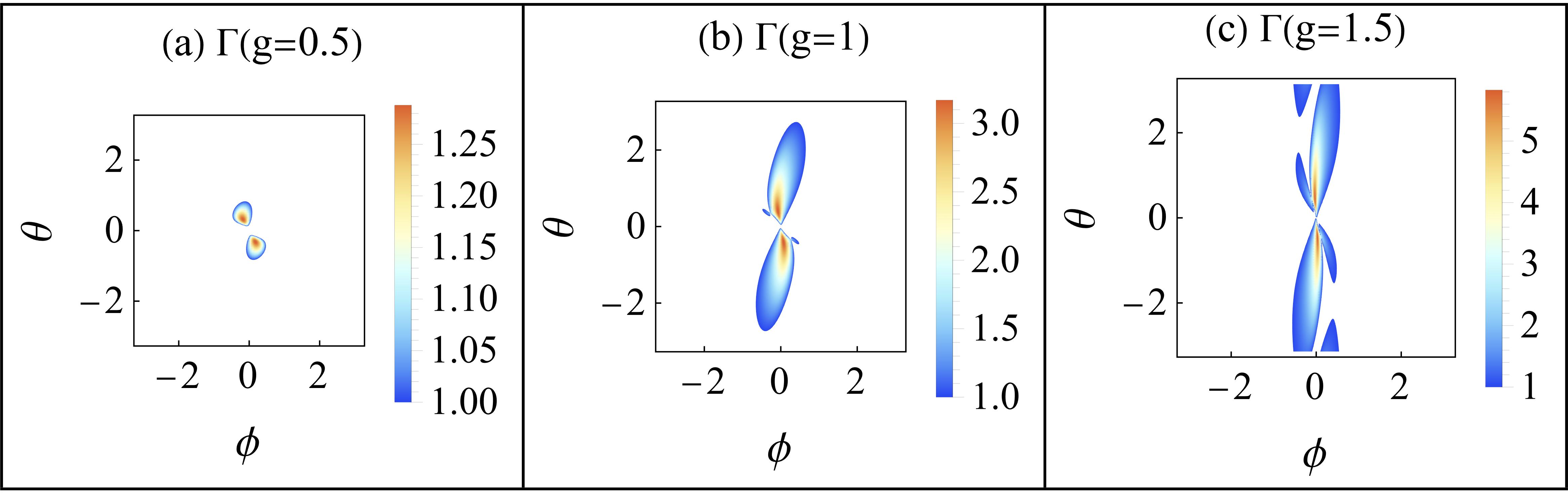}
\par\medskip 
\text{(A)} 
\vskip\baselineskip 
\includegraphics[width=1\linewidth, height=1.2in]{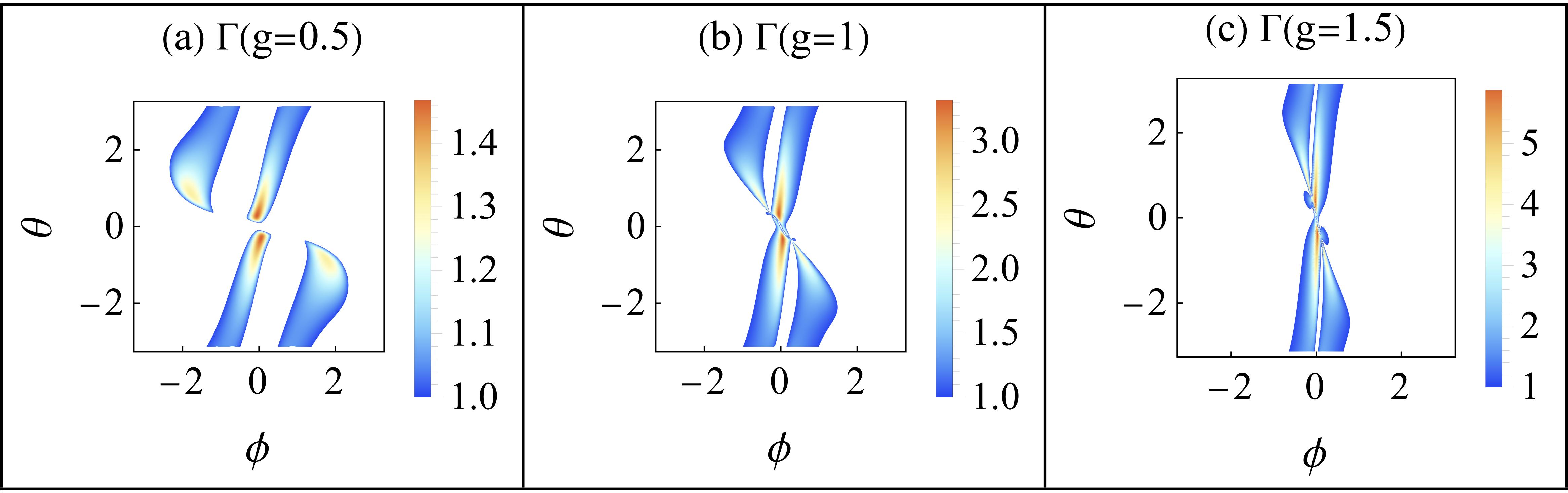}
\par\medskip 
\text{(B)}
\caption{The enhancement factor $\Gamma$, plotted as a function of phase angles $\theta$ and $\phi$ for (A) single-intensity detection and (B) homodyne detection. The plots are shown for OPA gains (a) $\text{g}=0.5$, (b) $\text{g}=1$, and (c) $\text{g}=1.5$, with the remaining parameters fixed at $\text{T}=0.80$, $\text{r}=0.5$ and $|\gamma|=1$.}
\label{Fig. 6}
\end{figure}

We further examine the effect of OPA gain \text{g} on the phase sensitivity of PR-DSU(1,1) interferometer and hence, the enhancement factor $\Gamma$ in our proposed scheme, as shown in Figs. ~\ref{Fig. 6}(A) (SID scheme) and ~\ref{Fig. 6}(B) (HD scheme). The enhancement factor $\Gamma$, exceeds unity (in the colored region), again indicating that our scheme achieves phase sensitivity beyond the SNL. Moreover, The results indicate that phase sensitivity and hence the enhancement factor significantly improve with increasing \text{g}.

\begin{figure}[htbp]
\centering
\includegraphics[width=1\linewidth, height=1.2in]{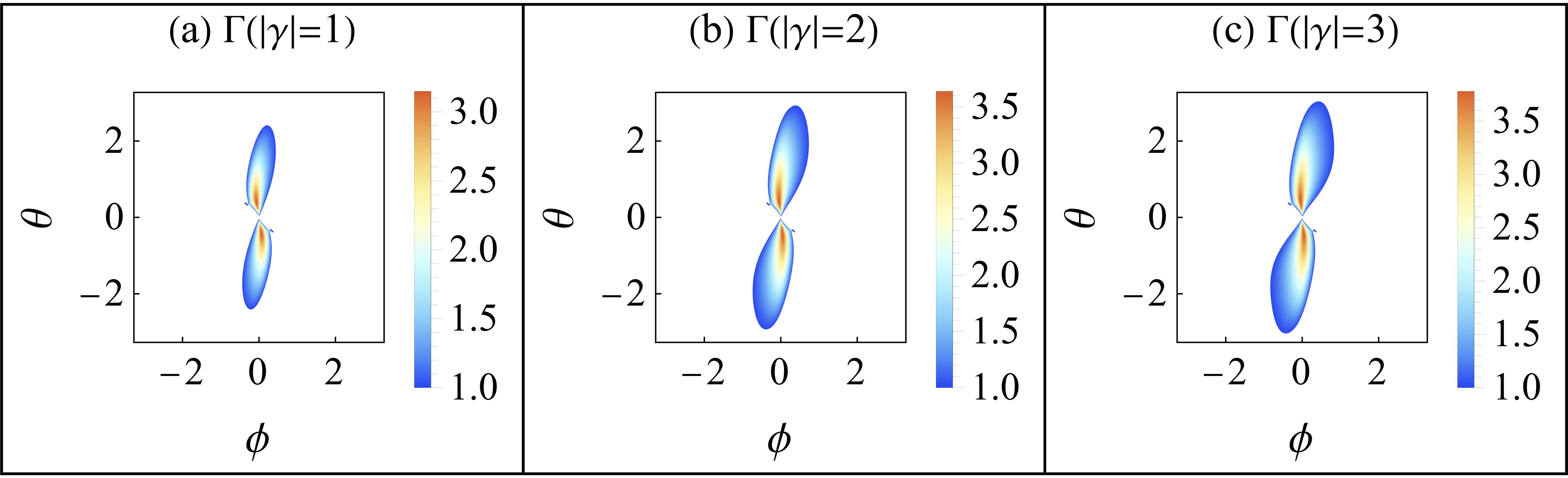}
\par\medskip 
\text{(A)}
\vskip\baselineskip 
\includegraphics[width=1\linewidth, height=1.2in]{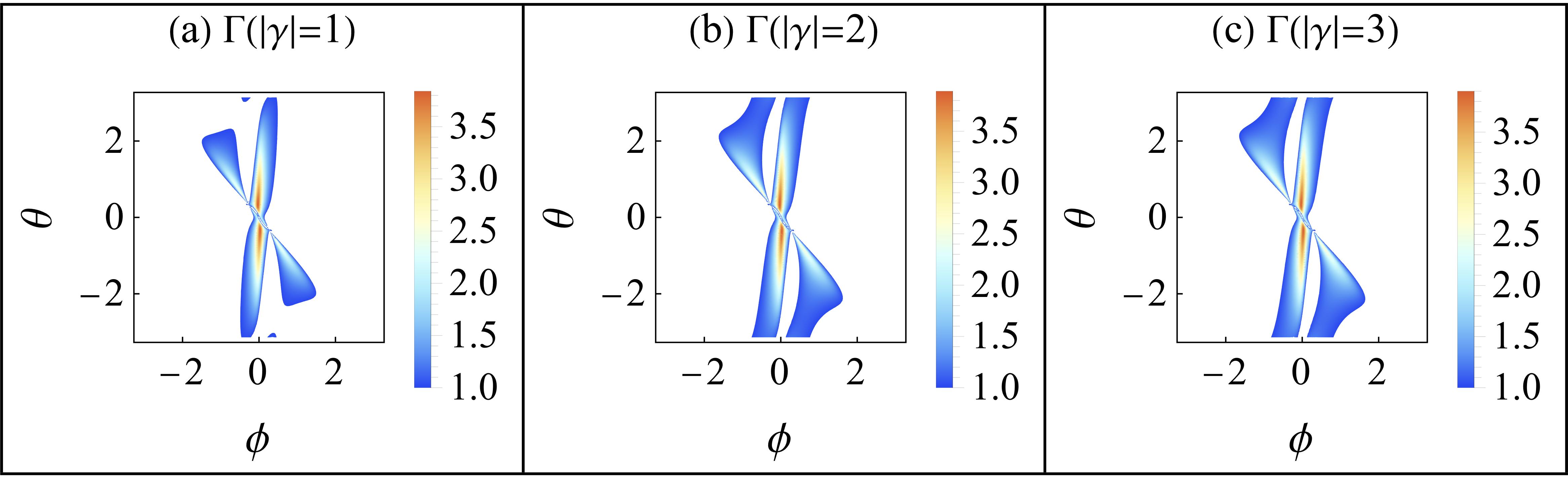}
\par\medskip 
\text{(B)}
\caption{The enhancement factor $\Gamma$, plotted as a function of phase angles $\theta$ and $\phi$ for (A) single-intensity detection and (B) homodyne detection. The plots are shown for displacement strengths (a)  $|\gamma|=1$,(b)  $|\gamma|=2$, (c) $|\gamma|=3$, with the remaining parameters fixed at $\text{g}=1$, $\text{r}=0.7$, $\text{T}=0.80$.}
\label{Fig. 7}
\end{figure}

Additionally, we explore the influence of displacement strength $\gamma$ on the enhancement factor $\Gamma$, as shown in Figs. ~\ref{Fig. 7}(A) (SID scheme) and ~\ref{Fig. 7}(B) (HD scheme). We observe an increase in phase sensitivity and hence the enhancement factor (exceeding unity in the colored region) with increasing $|\gamma|$ in the case of both detection schemes. However, this increase is more significant in the SID scheme. It is also important to note that in the absence of LDO, i.e., $|\gamma|=0$, the enhancement factor in the case of SID is found to be greater than unity only in the vicinity of $(\phi,\theta)=(0,0)$, but still smaller than that for the cases of $|\gamma|>0$. Moreover, the HD scheme will not be feasible for the case of $|\gamma|=0$ because it will cause the expectation value of the quadrature operator $\left\langle\hat{X}\right\rangle$ to be zero (see Eq. \ref{Eq. 15}, thereby making the sensitivity under the HD scheme to be uncertain.

\begin{figure}[htbp]
\centering
\includegraphics[width=1\linewidth, height=1.2in]{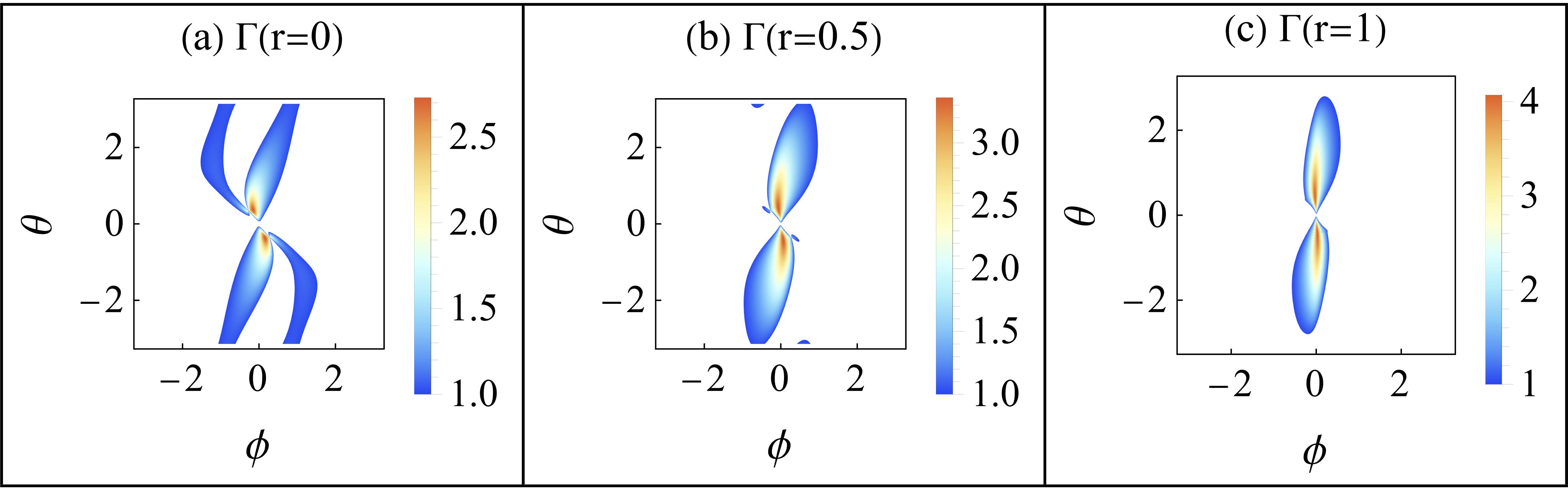}
\par\medskip 
\text{(A)} 
\vskip\baselineskip 
\includegraphics[width=1\linewidth, height=1.2in]{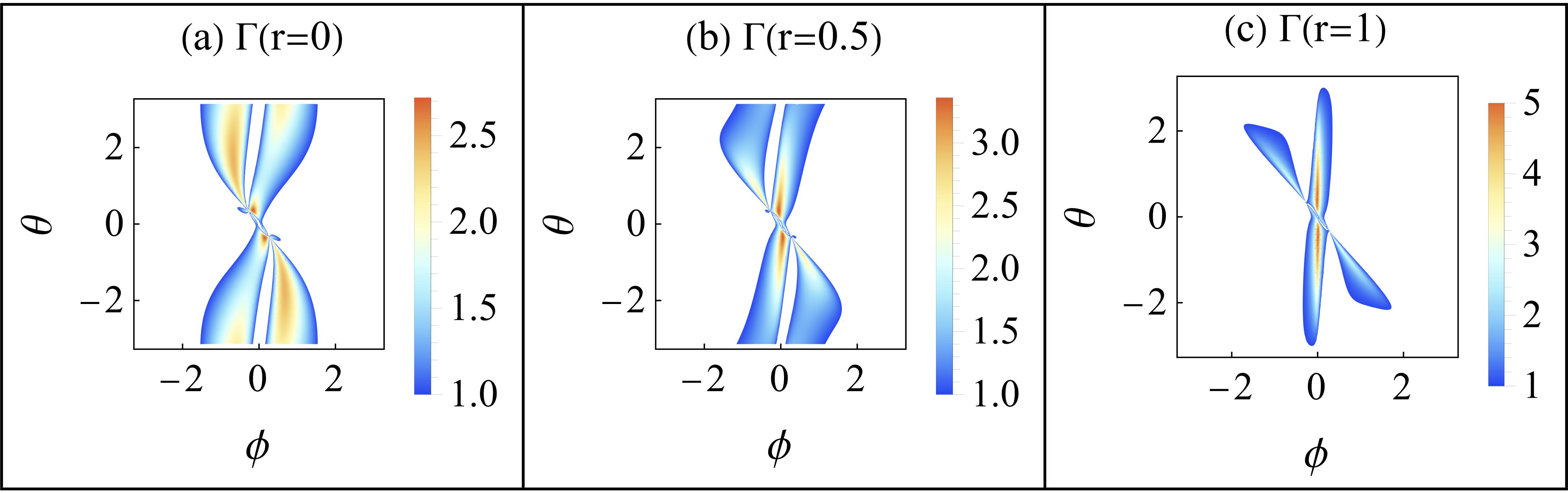}
\par\medskip 
\text{(B)}
\caption{The enhancement factor $\Gamma$, plotted as a function of phase angles $\theta$ and $\phi$ for (A) single-intensity detection and (B) homodyne detection. The plots are shown for different squeezing parameters (a)  $\text{r}=0$,(b)  $\text{r}=0.5$, (c) $\text{r}=1$, with the remaining parameters fixed at $\text{g}=1$, $\text{T}=0.80$, $|\gamma|=3$.}
\label{Fig. 8}
\end{figure}

 Finally, we examine the effect of squeezing parameter r on the enhancement factor $\Gamma$ in Figs. ~\ref{Fig. 8}(A) (SID scheme) and ~\ref{Fig. 8}(B) (HD scheme). We observe an increase in phase sensitivity and hence the enhancement factor (exceeding unity in the colored region) with increasing $\text{r}$ in the case of both detection schemes. We also notice that we can obtain the larger values of these enhancement factors for phase sensitivities (under both detection schemes) in the case of $\text{r}>0$ as compared to that at $\text{r}=0$. This highlights the advantage of squeezing ($\text{r} \ne 0$) due to reduced noise in the quadrature of the input light.

Comparing the performance of PR-DSU(1,1) interferometers under both detection schemes, we observe that although our scheme can achieve phase sensitivity beyond SNL, the HD scheme outperforms the SID scheme, which is obvious as the HD scheme is more sensitive than the simpler SID scheme. Additionally, we observe broader regions (values of $\theta$ and $\phi$) for $\Gamma > 1$ in the case of the HD scheme as compared to the SID scheme, further highlighting the superiority of the HD scheme over the SID scheme. Thus, we conclude that while the HD scheme involves a more complex setup than the SID scheme, it offers significantly greater precision, making it better suited for quantum-enhanced measurements.
\subsection{QCRB}
\begin{figure}[htbp]
\centering
\includegraphics[width=1\linewidth]{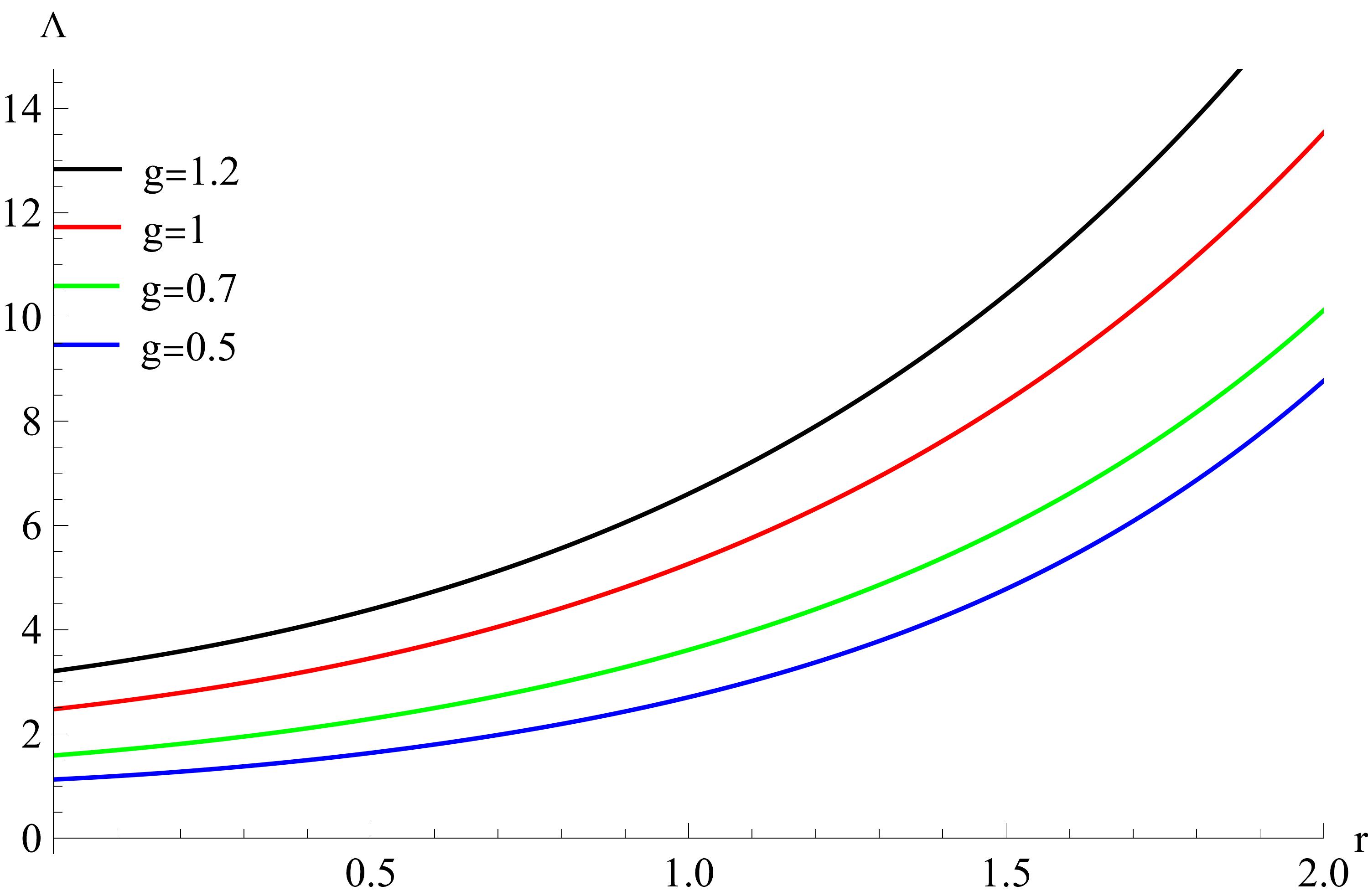}
\caption{\label{Fig._9} The enhancement factor $\Lambda$ with respect to squeezing parameter $\text{r}$ for different OPA gains, $\text{g}$= 0.5, 0.7, 1 and 1.2. Other parameters are $\text{T}=0.80$, $|\gamma|=1$, $\phi=0$, and $\theta=-\pi/8$. } 
\end{figure}

In order to compare the QCRB of our scheme with the SNL, we define
\begin{equation}\label{Eq. 31}
\Lambda=\frac{\Delta \phi_{\text{SNL}}}{\Delta\phi_{\text{QCRB}}^{\text{PR}}},
\end{equation}
where $\Lambda$ represents the enhancement factor for QCRB in our scheme compared to the SNL.
To analyze the effect of $\text{g}$, $\text{r}$, $\text{T}$ and $|\gamma|$ on $\Lambda$ for the optimal values of $\theta$ and $\phi$ for enhanced phase sensitivity, which are approximately the same in both detection schemes, we consider $\theta = -\pi/8$ and $\phi=0$. In Fig. ~\ref{Fig._9}, $\Lambda$ is plotted as a function of the squeezing parameter $\text{r}$ for different OPA gains $\text{g}$. We observe that similar to $\Gamma$, $\Lambda$ also exceeds unity, indicating that our scheme achieves QCRB beyond the SNL. Moreover, $\Lambda$ monotonically increases with r as well as \text{g}, indicating that QCRB is found to be significantly lower than the SNL and further decreases with increasing $\text{r}$ and $\text{g}$. Here also, we notice that similar to $\Gamma$, we can obtain the larger values of $\Lambda$ in the case of $\text{r}>0$ as compared to that at $\text{r}=0$.

\begin{figure}[htbp]
\centering
\includegraphics[width=1\linewidth]{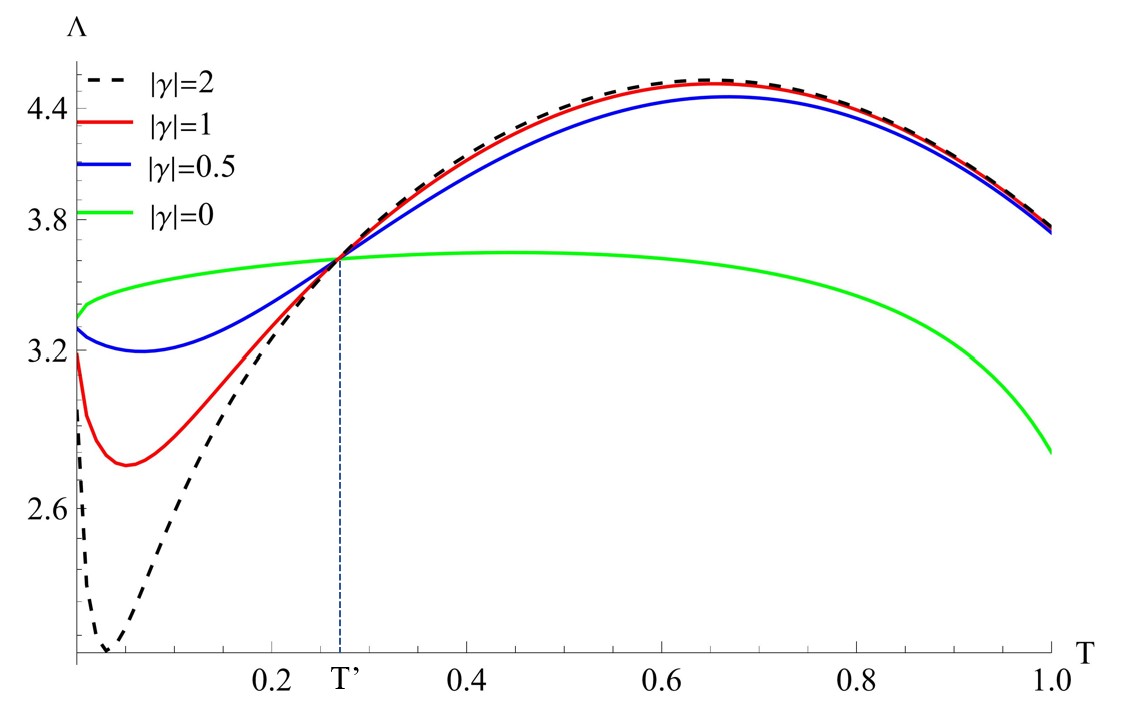}
\caption{\label{Fig._10} The enhancement factors $\Lambda$ with respect to transmission coefficient T for different displacement strengths, $|\gamma|$ = 0, 0.5, 1 and 2. Other parameters are $\text{g}$= 1.2, $\text{r} = 0.5$, $\phi=0$, and $\theta=-\pi/8$.} 
\end{figure}

Next, we examine the effects of transmission coefficient $\text{T}$ and the displacement strength $|\gamma|$ on QCRB in our scheme. To analyze it, we have plotted $\Lambda$ as a function of T with different values of LDO strength $|\gamma|$, as shown in Fig. ~\ref{Fig._10}. We observe that in the absence of LDO i.e., $|\gamma|=0$, $\Lambda$ monotonically increases with increasing T, which highlights the direct correlation of the increased number of recycled photons with enhanced sensitivity. However, this increase in $\Lambda$ is still relatively small, even at higher T, indicating that improvement in $\Lambda$ with recycled photons is limited due to the absence of LDO. On introducing LDO i.e., $|\gamma| \ne 0$, we observe a significant increment in the $\Lambda$ in the region T > T', which indicates that the photon recycling with LDO can enable us to achieve significant enhancement factor $\Lambda$ as compared to the case without photon recycling. However, this enhancement saturates after $|\gamma| = 1$. Moreover, we observe that in the region with T < T', $\Lambda$ significantly decreases with increasing $|\gamma|$.  Now, the value of T', after which introducing LDO becomes significantly beneficial, crucially depends on the values of OPA gain g and squeezing parameter r. One can easily verify that increasing g and decreasing r, will result in the value of T' shifting towards smaller values of T. In other words, we will get the benefit of LDO in making the enhancement factor much larger than the cases for T=0 over a large range of T by choosing larger g (such as 0.8 and onward) and smaller r (such as below 0.7). Therefore, with the proper choice of g and r and minimizing the photon loss in the feedback path, we get an optimal scenario where we can take advantage of photon recycling with LDO in order to achieve a significant value of $\Lambda$.

\begin{figure}[htbp]
\centering
\includegraphics[width=1\linewidth]{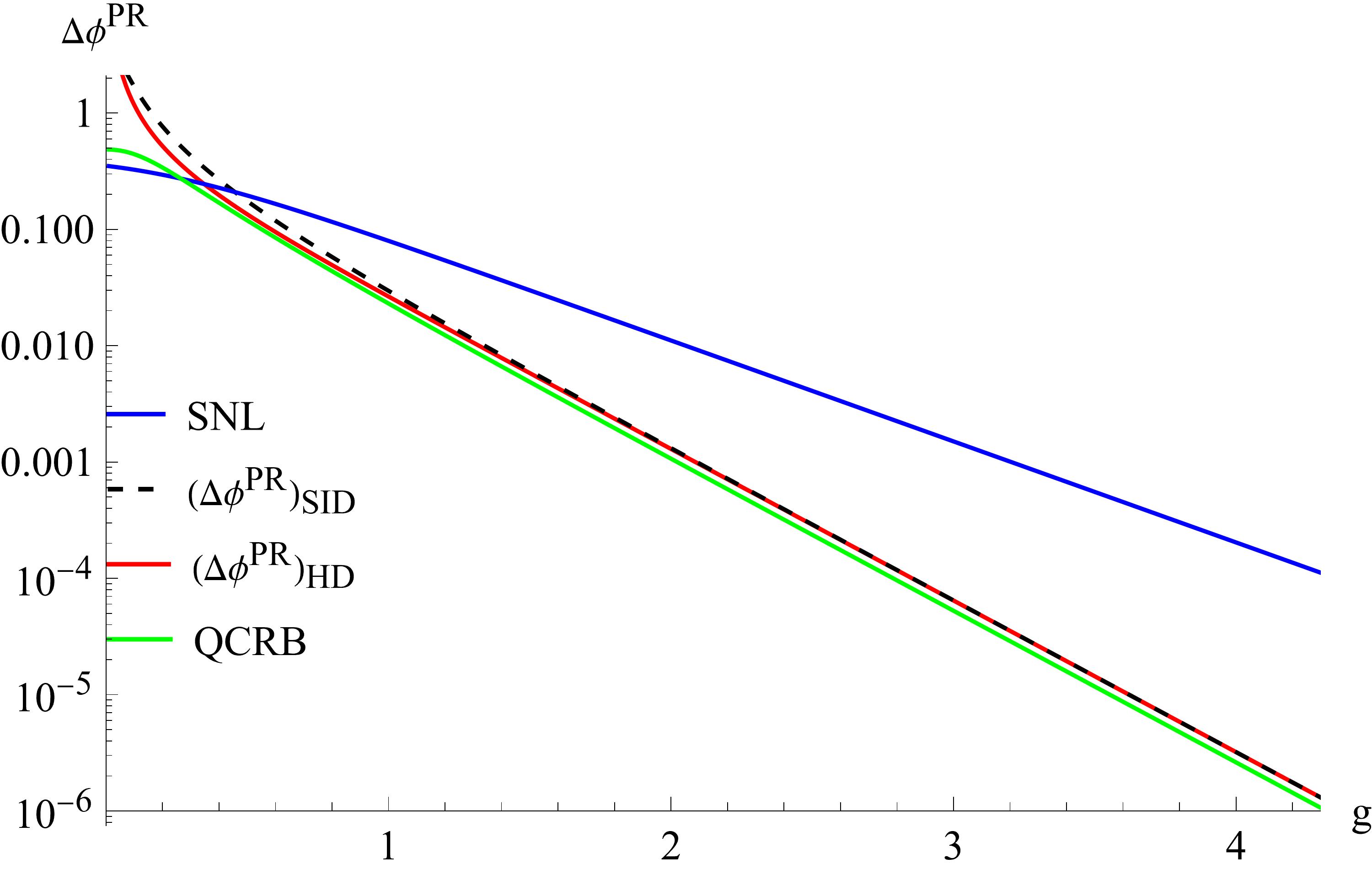}
\caption{\label{Fig._11} Phase sensitivity of the PR-DSU (1,1) interferometer under SID and HD detection schemes along with SNL and QCRB, plotted on a logarithmic scale, with respect to OPA gain $\text{g}$. Other parameters are $\text{T}=0.80$, $\text{r}=0.5$, $|\gamma|=1$, $\phi=0$, and $\theta=-\pi/8$.} 
\end{figure}

Now, we discuss the effect of increasing the gain of the OPAs on the performance of PR-DSU(1,1) interferometer. We have plotted the 
phase sensitivities of the PR-DSU(1,1) interferometer under both the detection schemes along with the SNL and QCRB on the logarithmic scale with respect to the OPA gain g as shown in Fig. ~\ref{Fig._11}. We observe that for the optimal values of $\theta$ and $\phi$, the total photon number (and hence SNL), QCRB, and the phase sensitivities under both detection schemes are not saturated and decrease exponentially on increasing g. From the experimental point of view (where the maximum achievable squeezing parameter for OPA is around $g \approx 1.2$ \cite{SCHNABEL20171,Schonbeck:18,PhysRevLett.117.110801}), we can take advantage of the improved performance of this PR-DSU(1,1) interferometer, as it is found to be improving with increasing g even with the experimentally achievable conditions. Moreover, we observe the superiority of the HD scheme over the SID scheme for the values of g in the range from 0.5 to 1.5, as we have already seen in Fig. 4. However, phase sensitivities under both detection schemes concide for the larger values of g. 

\section{Conclusions} \label{Section 7}

In summary, we propose a novel approach to enhance phase estimation in the DSU(1,1) interferometer through photon recycling. Starting with a DSU(1,1) interferometer having a vacuum state at port ``a'' and a squeezed vacuum state (with squeezing parameter r) at port ``b'', we introduced a phase shift $\phi$ in the arm ``b''. This setup is modified by re-injecting the output mode ``a'' back into the input mode after a phase shift, $\theta$, and photon loss is characterized by $\sqrt{1-\text{T}}$. We examined the phase sensitivity of the PR-DSU(1,1) interferometer $\Delta\phi^{\text{PR}}$ under both single-intensity detection and homodyne detection schemes. To provide the ultimate theoretical limit, we calculated the quantum Cram\'{e}r-Rao bound (QCRB), denoted as $\Delta\phi_{\text{QCRB}}^{\text{PR.}}$. In order to analyze the improvement in performance induced by photon recycling, we have compared the performance of the PR-DSU(1,1) interferometer with the conventional DSU(1,1) interferometer by introducing two enhancement factors, $\Sigma$ and $\Xi$, where $\Sigma=\Delta\phi^{\text{Conv.}}/\Delta\phi^{\text{PR}}$ and $\Xi=\Delta\phi_{\text{QCRB}}^{\text{Conv.}}/\Delta\phi_{\text{QCRB}}^{\text{PR.}}$ represent the enhancement factors for phase sensitivity and QCRB of PR-DSU(1,1) interferometer relative to DSU(1,1) interferometer, respectively. In addition, for each detection scheme, we calculated the enhancement factor for phase sensitivity of the PR-DSU(1,1) interferometer relative to SNL, defined as $\Gamma=\Delta \phi_{\text{SNL}}/\Delta\phi^{\text{PR}}$, and compared the results. Similarly, we calculated the enhancement factor for QCRB of the PR-DSU(1,1) interferometer relative to SNL, defined as $\Lambda=\Delta \phi_{\text{SNL}}/\Delta\phi_{\text{QCRB}}^{\text{PR.}}$.

We have plotted $\Sigma$ and $\Xi$ as a function of T, as shown in Figs. \ref{Fig._3} and \ref{Fig._4}, respectively. The values of these enhancement factors exceeding unity clearly indicate that the PR-DSU(1,1) interferometer can possess phase sensitivity and QCRB beyond those in the case of the conventional DSU(1,1) interferometer. This highlights the importance of photon recycling in enhancing the sensitivity of the interferometer.
Next, we plotted $\Gamma$  as the function of phases $\phi$ and $\theta$ for varying values of T, \text{g}, $|\gamma|$, and \text{r} (see Figs. \ref{Fig. 5}(A), ~\ref{Fig. 6}(A), ~\ref{Fig. 7}(A) and ~\ref{Fig. 8}(A) (SID scheme) and Fig. \ref{Fig. 5}(B), \ref{Fig. 6}(B), ~\ref{Fig. 7}(B) and ~\ref{Fig. 8}(B) (HD scheme), respectively). The values of the enhancement factor exceeding unity (in the colored region) indicate that our scheme can achieve phase sensitivity beyond the SNL. Moreover, it increases with an increase in T, \text{g}, $|\gamma|$, and \text{r}.
On comparing the phase sensitivity of the PR-DSU(1,1) interferometer under both SID and HD schemes, we observe that the HD scheme outperforms the SID scheme. The HD scheme also exhibits a broader region of $\theta$ and $\phi$ for enhanced phase sensitivity beyond SNL, further demonstrating its superiority over the SID scheme.  Similarly, we have plotted $\Lambda$  as the function of  \text{r} for varying values of \text{g} (see Fig. \ref{Fig._9}) and observed that the value of $\Lambda$ exceeds unity and increases monotonically with increasing \text{g} and \text{r}. By plotting $\Lambda$ as the function of  \text{T} for varying values of $|\gamma|$ (see Fig. \ref{Fig._10}), we observed a significant improvement due to recycled photons when accompanied by LDO, depending upon the proper choice of \text{g}, \text{r}, and the amount of photon loss in the feedback arm.

Therefore, our findings show that this modified scheme has phase sensitivity and QCRB beyond the SNL as well as beyond those in the case of conventional DSU(1,1) interferometer, offering a novel approach to enhancing phase sensitivity via photon recycling.

\begin{acknowledgments}
TK acknowledges MoE, Government of India for the Prime Minister's Research Fellowship. 
AK acknowledges UGC for the UGC Research Fellowship.
AKP and DKM acknowledge financial support from the Science \& Engineering Research Board (SERB), New Delhi for the CRG Grant (CRG/2021/005917). DKM acknowledges the Chanakya Doctoral Fellowship grant (I-HUB/DF/2022-23/04) and Incentive Grant under the Institution of Eminence (IoE), Banaras Hindu University, Varanasi, India.
\end{acknowledgments}

\section*{Data Availability Statement}
No data were generated or analyzed in the presented research.

\nocite{*}

\bibliographystyle{apsrev4-1}  

\bibliography{aipsamp}

\end{document}